\documentclass[aps,prb,superscriptaddress]{revtex4-2}

\usepackage[T1]{fontenc}
\usepackage{lmodern}
\usepackage{amsmath,amssymb,amsfonts}
\usepackage{bm}
\usepackage{graphicx}
\usepackage{booktabs}
\usepackage{microtype}
\usepackage{xcolor}
\usepackage{hyperref}
\hypersetup{hidelinks}
\usepackage{cleveref}
\usepackage{enumitem}
\usepackage{mathtools}
\usepackage{tcolorbox}
\usepackage{tikz}
\usetikzlibrary{positioning}
\newcommand{\dd}{\mathrm{d}}
\newcommand{\kb}{k_{\mathrm B}}

\newcommand{\rr}{\mathbf{r}}
\newcommand{\x}{\mathbf{x}}
\newcommand{\q}{\mathbf{q}}
\newcommand{\uu}{\mathbf{u}}
\newcommand{\hh}{\mathbf{h}}

\newcommand{\Pc}{P_{\mathrm c}}
\newcommand{\Pfour}{P_4}
\newcommand{\Pfive}{P_5}
\newcommand{\eps}{\epsilon}
\newcommand{\dc}{d_c}

\newtcolorbox{physicalmessage}{
  title=Physical message,
  colback=gray!5,
  colframe=black!60,
  fonttitle=\bfseries
}

\newtcolorbox{keyresult}{
  title=Key result,
  colback=gray!5,
  colframe=black!60,
  fonttitle=\bfseries
}

\newtcolorbox{openproblem}{
  title=Open question,
  colback=gray!5,
  colframe=black!60,
  fonttitle=\bfseries
}

\begin{document}

\title{\textbf{Quenched disorder in polymerized membranes: from anomalous elasticity to a glassy phase}}
 
\author{D. Mouhanna} 
\email{mouhanna@lptmc.jussieu.fr}
\affiliation{Sorbonne Universit\'e, CNRS, Laboratoire de Physique Th\'eorique de la Mati\`ere Condens\'ee, LPTMC, F-75005 Paris, France}

\begin{abstract}
Polymerized membranes provide a particularly rich class  of fluctuating
extended systems in which geometry, elasticity and statistical fluctuations 
are intimately coupled and generate nontrivial long-distance physics.  
In their disorder-free flat phase, the nonlinear coupling between in-plane phonons and out-of-plane
flexural modes produces anomalous elasticity: the effective bending
rigidity grows algebraically at long wavelength while the in-plane
elastic moduli soften.  Quenched disorder enriches this picture
considerably.  In particular, random metric and random curvature
couple directly to the elastic and bending sectors and generate a
competition between thermal and disorder-induced fluctuations.

We review the development of this problem, first discussing perturbative and nonperturbative field-theoretical approaches to pure polymerized membranes and their anomalous elasticity, and then extending the discussion to the effects of quenched disorder. Early one-loop analyses found the disorder-free fixed point, 
$\Pfour$, and a zero-temperature disorder-dominated fixed point,
$\Pfive$, but no finite-temperature transition to a glassy phase.
However, a  nonperturbative renormalization-group analysis subsequently revealed
a third fixed point, $\Pc$, at finite temperature and finite disorder.
Within this description $\Pc$ controls a continuous wrinkling
transition between a thermally dominated flat phase governed by
$\Pfour$ and a low-temperature disorder-dominated phase, interpreted within this framework as a glassy phase, governed by $\Pfive$.

Higher-order perturbative calculations recover a $\Pc$-like fixed
point and yield anomalous dimensions whose $\epsilon$ expansions are
remarkably close to those obtained by reexpanding the nonperturbative
flow.

A particularly important aspect of this scenario is its connection with the wrinkling transition first observed several decades ago in partially polymerized lipid membranes.  The three experimentally observed scaling regimes of the
height fluctuations can be associated with $\Pfour$, $\Pc$, and $\Pfive$, and the corresponding theoretical power-law exponents show good overall quantitative agreement with experiment, with particularly striking agreement at the critical point.  We finally discuss the status of this glassy scenario, the remaining discrepancy
between fixed-point stability properties in finite-order perturbation
theory and in the nonperturbative approach, and possible extensions to
defective graphene and related two-dimensional crystalline materials.
\end{abstract}

\maketitle

\section{Introduction}
\label{sec:introduction}

Membrane physics appears in a broad variety of contexts \cite{proceedings89,bowick01}. In soft and
biological matter, lipid bilayers provide paradigmatic examples of
fluid membranes, while partially polymerized membranes and membranes
coupled to an elastic cytoskeleton possess additional in-plane elastic
degrees of freedom. In condensed-matter physics, graphene and related
atomically thin crystalline materials constitute genuine
two-dimensional elastic membranes whose remarkable mechanical
properties are strongly influenced by flexural fluctuations
\cite{proceedings89,bowick01}.
These systems illustrate a fundamental peculiarity of membrane
physics: thermal fluctuations do not merely perturb a rigid
background, but can qualitatively modify the elastic properties of the
membrane at large distances \cite{nelson87}.

We begin by briefly introducing the geometrical and field-theoretical
framework used to describe polymerized membranes. 
A $D$-dimensional membrane
embedded in a $d$-dimensional Euclidean space is described by an
embedding field
\begin{equation}
  \rr : \x\in\mathbb{R}^{D}
  \longrightarrow
  \rr(\x)\in\mathbb{R}^{d},
  \label{eq:embedding}
\end{equation}
where $\x=(x_1,\ldots,x_D)$ denotes a set of internal coordinates.
The physical case of an ordinary membrane corresponds to $D=2$ and
$d=3$.  The large-distance morphology of such an object may be
controlled by thermal fluctuations, elastic interactions, quenched
disorder, or by a competition among these ingredients.

A first fundamental distinction is that between \emph{fluid} and
\emph{polymerized}, or crystalline, membranes
\cite{nelson87,proceedings89,bowick01}.  In a fluid membrane the
constituents can diffuse within the membrane plane, so that the
connectivity is not fixed and the static shear modulus vanishes.  The
dominant long-wavelength elastic cost is therefore associated with
curvature.  In a polymerized membrane, by contrast, the connectivity
between constituents is fixed.  In-plane deformations then cost
elastic energy, and the membrane possesses phonon modes in addition to
out-of-plane flexural fluctuations.  This apparently elementary
difference changes the infrared physics qualitatively.

The flat phase of a polymerized membrane is one of the most striking
examples. Indeed,  polymerized membranes can sustain an ordered flat phase at low temperature \cite{nelson87,aronovitz88,guitter89,aronovitz89,coquand19b}.
This result is far from being trivial. If one retains only harmonic bending fluctuations, the
fluctuations of the local normal diverge logarithmically in two
dimensions and long-range orientational order appears impossible.
The resolution of this apparent paradox is provided by the nonlinear geometrical coupling
between in-plane phonons and out-of-plane flexural modes.  This
coupling induces effective long-range interactions and produces
anomalous, scale-dependent elastic constants
\cite{nelson87,aronovitz88,aronovitz89,guitter88,guitter89}.  In the flat phase the
effective bending rigidity behaves as
\begin{equation}
  \kappa_{\mathrm{eff}}(q)
  \underset{q\rightarrow0}{\sim}
  q^{-\eta},
  \label{eq:intro-kappa}
\end{equation}
where $\eta$ is a universal anomalous dimension.  At the same time,
the in-plane elastic moduli vanish algebraically at long wavelength.
Thermal fluctuations therefore stiffen the membrane against bending
while softening its in-plane elastic response
\cite{nelson87,aronovitz88,aronovitz89,guitter88,guitter89}.

This anomalous elasticity has been studied by a variety of
field-theoretical methods: the $\epsilon=4-D$ expansion,
large-$d$ expansions, the self-consistent screening
approximation (SCSA), and nonperturbative renormalization-group
(NPRG) methods
\cite{aronovitz88,guitter88,guitter89,aronovitz89,ledoussal92,kownacki09}.  A consistent picture has emerged in which a
nontrivial infrared fixed point, denoted below by $\Pfour$, controls
the flat phase.  In this context the difficulty is primarily quantitative: the physical case, $D=2$ and $d=3$, lies far from both perturbative limits, $D=4$ and large $d$. This has motivated the development of NPRG 
approaches to polymerized membranes \cite{kownacki09,hasselmann11}.

The problem becomes even richer in the presence of quenched disorder, see \cite{proceedings89,bowick01} for reviews.  Real
membranes and crystalline monolayers are never perfectly homogeneous.
Defects, vacancies, impurities, heterogeneous polymerization, local
variations of bonding, and structural rearrangements can produce
frozen distortions of the preferred metric and of the preferred
curvature \cite{morse92a,morse92b}.  At the level of the elastic continuum theory, these
effects are naturally represented by \emph{metric disorder} and
\emph{curvature disorder}.  They couple to different geometrical
sectors of the membrane and compete with thermal fluctuations.

The motivation for studying this competition has two complementary
origins.  The first is the physics of graphene and related
two-dimensional crystalline materials \cite{novoselov04,castroneto09,katsnelson12,amorim16,akinwande17}.  Defects are unavoidable and
can affect mechanical, thermal, transport, and electronic properties of these systems.  The elastic membrane description has been remarkably successful
for the long-distance properties of pristine graphene, suggesting
that it is a natural starting point for investigating the effect
of static structural disorder
\cite{castroneto09,amorim16,katsnelson13,gornyi15}.

The second motivation comes from an older set of experiments on
partially polymerized phospholipid membranes.  By varying the degree of polymerization, Mutz, Bensimon, and Brienne  observed a reversible transformation from relatively smooth vesicles to strongly wrinkled structures \cite{mutz91}. Chaieb and collaborators subsequently characterized the associated height fluctuations quantitatively \cite{chaieb06,chaieb08}.  The low-polymerization state  was conjectured to be related to a membrane analogue of a glassy transition, in which frozen geometrical fluctuations play the role usually played by frozen random degrees of freedom in spin
glasses \cite{mutz91,chaieb06,chaieb08}.

Early theoretical work produced a puzzling situation.
Nelson and Radzihovsky found that weak metric disorder is irrelevant
at finite temperature close to $D=4$, although disorder can become
important as the temperature tends to zero
\cite{nelson91,radzihovsky91b}.
Morse, Lubensky, and Grest \cite{morse92a} and Morse and Lubensky \cite{morse92b}  then included
curvature disorder.

Their one-loop analysis identified a new
zero-temperature disordered fixed point, denoted $\Pfive$, but found
it marginally unstable with respect to temperature
\cite{morse92a,morse92b}.  Thus the early weak
coupling picture contained an anomalous disorder-controlled regime but
no finite-temperature critical point separating a thermal flat phase
from a genuine low-temperature glassy phase.

A qualitatively different picture  emerged from the NPRG analysis of
the same metric--curvature disorder problem
\cite{coquand18}.  Besides the familiar
finite-temperature disorder-free fixed point $\Pfour$ and the
zero-temperature disordered fixed point $\Pfive$, the NPRG flow revealed  a
third fixed point $\Pc$ at finite temperature and finite disorder.
The point $\Pc$ has one relevant direction and controls a continuous
transition between two basins of attraction.  On the high-temperature
side, trajectories flow to $\Pfour$ and disorder becomes
asymptotically irrelevant.  On the low-temperature side, trajectories
flow to $\Pfive$, which is stable within the NPRG and controls an
entire disorder-dominated phase.  Within this picture the latter
constitutes an entire glassy phase rather than a transient crossover
regime.

Two subsequent developments are particularly important.  First, the
scaling exponents associated with $\Pfour$, $\Pc$, and $\Pfive$ were
compared with the height-fluctuation measurements of partially
polymerized lipid membranes.  The three predicted universal regimes show good overall quantitative agreement with the smooth, critical, and wrinkled morphologies observed experimentally, with particularly close agreement in the critical regime \cite{coquand20}.  Second, perturbation theory
was pushed beyond one loop.  A two-loop analysis revealed a
finite-temperature, finite-disorder $\Pc$-like fixed point \cite{coquand20b} and a
three-loop calculation confirmed its existence while correcting the
two-loop RG functions
\cite{metayer22d}.
The anomalous dimensions obtained perturbatively are structurally and
numerically very close to the $\epsilon$ expansion of the NPRG
results, even though a discrepancy remains in the finite-order
stability properties of $\Pc$ and $\Pfive$.  The agreement between methods based on very different approximation schemes provides strong support for the existence of corresponding fixed-point solutions with closely related anomalous dimensions.

The purpose of this review is to organize these results into a single
physical picture.  We first recall the geometrical and elastic
description of fluid and polymerized membranes and the anomalous
elasticity of the disorder-free flat phase.  We then introduce metric
and curvature disorder, review the early perturbative scenario,
describe the NPRG fixed-point structure, and discuss the higher-loop
perturbative calculations.  The connection with partially polymerized
lipid membranes is treated separately because it provides a direct
experimental interpretation of the three fixed points.  We conclude
with the open theoretical and numerical issues and with the possible
relevance of this scenario for defective two-dimensional materials.

\begin{physicalmessage}
The central point of this review is that quenched disorder does not
merely renormalize the elastic properties of the thermal flat phase.
The combined metric--curvature disorder problem supports a richer RG
topology involving the thermal fixed point $\Pfour$, a critical
wrinkling fixed point $\Pc$, and a disorder-dominated fixed point
$\Pfive$ \cite{coquand18,coquand20,coquand20b,metayer22d}.
\end{physicalmessage}

\section{From fluid to polymerized membranes}
\label{sec:fluid-polymerized}

\subsection{Geometry and bending energy of an embedded membrane}

Let us first recall the geometrical quantities entering the continuum
description of a membrane. The tangent vectors associated with the
embedding Eq.~\eqref{eq:embedding} are
\begin{equation}
  \mathbf{e}_a(\x)=\partial_a\rr(\x),
  \qquad a=1,\ldots,D,
\end{equation}
and define the induced metric
\begin{equation}
  g_{ab}=\partial_a\rr\cdot\partial_b\rr.
  \label{eq:metric}
\end{equation}
For a two-dimensional membrane embedded in three dimensions one may
introduce the unit normal ${\bf n}=\displaystyle {{\bf e}_1\times {\bf e}_2\over \vert {\bf e}_1\times {\bf e}_2 \vert }$
 and the curvature tensor:
 \begin{equation}
 K_{ab}=-{\bf n}.\, \partial_b{\bf e}_a={\bf e}_a.\, \partial_b{\bf n}. 
 \end{equation}
If $K_1$ and $K_2$ are its principal curvatures, the mean
and Gaussian curvatures are
\begin{equation}
  H=\frac{1}{2}(K_1+K_2),
  \qquad
  K=K_1K_2.
\end{equation}

For a tensionless membrane with up--down symmetry and vanishing
spontaneous curvature, the local bending energy contains both
mean- and Gaussian-curvature contributions \cite{helfrich73},
\begin{equation}
  F_{\rm bend}
  =
  \int_{\mathcal M} dA\,
  \left[
    \frac{\kappa}{2}(2H)^2
    +\bar\kappa K
  \right],
\end{equation}
where $\kappa$ and $\bar\kappa$ are the bending and Gaussian rigidities,
respectively.

For a closed membrane of fixed topology, the Gauss--Bonnet theorem
implies
\begin{equation}
  \int_{\mathcal M} K\,dA
  =
  2\pi\chi(\mathcal M),
\end{equation}
where  $\chi(\mathcal M)$ is  the Euler characteristic of  the manifold  ${\mathcal M} $, so that the Gaussian-curvature contribution is a topological constant
and does not affect bulk shape fluctuations. For an open membrane,
which is the situation most commonly relevant for extended polymerized
or crystalline sheets, the corresponding relation contains an
additional boundary contribution,
\begin{equation}
  \int_{\mathcal M} K\,dA
  +
  \oint_{\partial\mathcal M} k_g\,ds
  =
  2\pi\chi(\mathcal M),
\end{equation}
where $k_g$ is the geodesic curvature of the boundary. Thus, for an
open membrane the integral of the Gaussian curvature is not by itself
topologically fixed and may affect boundary physics. In the following,
however, we are interested in the bulk long-wavelength properties of
extended membranes and neglect boundary effects. We therefore omit the
Gaussian-curvature contribution and retain the mean-curvature term,
\begin{equation}
  F_{\rm bend}
  =
  \frac{\kappa}{2}
  \int d^2x\,\sqrt{g}\,(2H)^2 .
\end{equation}

Close to a flat configuration, one may use a Monge parametrization,
$\mathbf r(x,y)=(x,y,h(x,y))$. Expanding the bending energy for small
slopes gives, at harmonic order,
\begin{equation}
  F_{\rm bend}
  \simeq
  \frac{\kappa}{2}
  \int d^2x\,(\Delta h)^2 .
  \label{eq:harmonic_bending}
\end{equation}

\subsection{Fluid membranes}

In a fluid membrane, the constituent molecules are free to diffuse
within the membrane plane.  The absence of fixed connectivity implies
a vanishing static shear modulus, so that the bending sector dominates
the long-wavelength elastic description \cite{bowick01}.

The harmonic propagator obtained from
Eq.~\eqref{eq:harmonic_bending} is
\begin{equation}
  \left\langle h(\q)h(-\q)\right\rangle
  =
  \frac{\kb T}{\kappa q^4}.
  \label{eq:fluid-propagator}
\end{equation}
For small slopes the fluctuations of the local normal are controlled
by $\nabla h$, and in $D=2$ one has 
\begin{equation}
  \langle\theta^2\rangle
  \sim
  \kb T\int\frac{\dd^2q}{(2\pi)^2}
  \frac{1}{\kappa q^2}
  \sim
  \frac{\kb T}{\kappa}\ln\frac{1}{qa} \underset{q \to 0} \longrightarrow   \infty.
  \label{eq:normal-fluid}
\end{equation}
The infrared divergence indicates the absence of long-range
orientational order of the normals in the harmonic theory.
Anharmonic curvature fluctuations do not stabilize such an order;
rather, they soften the effective rigidity of a fluid membrane. 
For a two-dimensional fluid membrane embedded in $d$ dimensions, the one-loop renormalized bending rigidity reads \cite{santiago05}
\begin{equation}
  \kappa_R(\q)=\kappa - \frac{d\kb T}{4\pi}  \ln\frac{1}{qa} 
\end{equation}
which makes the divergence in (\ref{eq:normal-fluid}) worse. 

This behavior strongly contrasts with that of polymerized membranes, for which in-plane elasticity reverses the trend and produces an infrared enhancement of
the effective rigidity \cite{nelson87}.

\subsection{Polymerized membranes}

A polymerized membrane is characterized by a fixed connectivity between
its constituents. It therefore resists not only bending but also stretching
and shear \cite{nelson87,bowick01}. Examples range from polymerized or
tethered membranes in soft matter to crystalline monolayers such as
graphene.

In the flat phase, rotational symmetry is spontaneously broken and the
ordered state can be characterized by a nonvanishing expectation value
of the tangent vectors,
\begin{equation}
  \left\langle \partial_a \rr(\x) \right\rangle
  =
  \zeta\,\mathbf e_a,
  \qquad a=1,\ldots,D,
  \label{eq:flat-order-parameter}
\end{equation}
where the $\mathbf e_a$ form an orthonormal set of $D$ vectors in the
embedding space. The stretching parameter $\zeta$ measures the amplitude of the
flat-phase order parameter. Up to an arbitrary uniform translation,
Eq.~\eqref{eq:flat-order-parameter} is equivalently written as
\begin{equation}
  \left\langle \rr(\x)\right\rangle
  =
  \zeta\sum_{a=1}^{D}x_a\,\mathbf e_a .
  \label{eq:flat-reference}
\end{equation}
It is therefore natural to define the flat reference configuration as
\begin{equation}
  \rr_0(\x)
  \equiv
  \left\langle \rr(\x)\right\rangle ,
  \qquad
  \partial_a\rr_0
  =
  \zeta\,\mathbf e_a .
  \label{eq:r0-definition}
\end{equation}
The corresponding reference metric is thus
\begin{equation}
  g^{0}_{ab}
  =
  \partial_a\rr_0\cdot\partial_b\rr_0
  =
  \zeta^2\delta_{ab}.
  \label{eq:reference-metric}
\end{equation}

Fluctuations around this ordered configuration naturally separate into
in-plane phonon modes $\uu$ and out-of-plane flexural modes $\hh$,
\begin{equation}
  \rr(\x)
  =
  \rr_0(\x)+\uu(\x)+\hh(\x),
  \label{eq:decomposition}
\end{equation}
where $\uu$ belongs to the $D$-dimensional tangent subspace spanned by
the $\mathbf e_a$, whereas $\hh$ belongs to the orthogonal subspace and
has $d_c=d-D$ components.

The strain tensor measures the deviation of the induced metric from its
reference value,
\begin{equation}
  u_{ab}
  =
  \frac{1}{2}
  \left[
    \partial_a\rr\cdot\partial_b\rr
    -
    \partial_a\rr_0\cdot\partial_b\rr_0
  \right].
  \label{eq:strain-geometric}
\end{equation}
For the long-wavelength description of the flat phase, and setting
$\zeta=1$ for simplicity, this reduces to
\begin{equation}
  u_{ab}
  \simeq
  \frac{1}{2}
  \left(
    \partial_a u_b+\partial_b u_a
    +
    \partial_a\hh\cdot\partial_b\hh
  \right),
  \label{eq:strain}
\end{equation}
where terms quadratic in the in-plane displacement have been neglected,
while the geometrically required nonlinear coupling between the phonon
and flexural sectors has been retained.

The effective elastic free energy is \cite{nelson87,aronovitz88,aronovitz89}
\begin{equation}
  F[\uu,\hh]
  =
  \int\dd^Dx
  \left[
    \frac{\kappa}{2}(\Delta\hh)^2
    +
    \frac{\lambda}{2}u_{aa}^2
    +
    \mu u_{ab}u_{ab}
  \right],
  \label{eq:pure-action}
\end{equation}
where $\lambda$ and $\mu$ are the Lam\'e coefficients.  Mechanical
stability requires
\begin{equation}
  \kappa>0,\qquad
  \mu>0,\qquad
  \lambda+\frac{2}{D}\mu>0.
\end{equation}

The crucial element of Eq.~\eqref{eq:pure-action} is the nonlinear
phonon--flexural coupling contained in Eq.~\eqref{eq:strain}.  Once
the in-plane degrees of freedom are integrated out, it generates
long-range effective interactions between curvature fluctuations.
These interactions suppress the infrared divergence that destabilizes
the orientational order of a fluid membrane.  The flat phase of a
polymerized membrane is therefore a genuine ordered phase with
nontrivial scale-dependent elastic parameters
\cite{nelson87,aronovitz88,aronovitz89,guitter89}.

The resulting mechanism is unusual from the viewpoint of conventional
two-dimensional statistical mechanics. Although the membrane is a
two-dimensional system, the effective interactions between its
orientational degrees of freedom become sufficiently long ranged to
allow long-range order of the normals. There is therefore no
contradiction with the Mermin--Wagner theorem, because integrating out
the in-plane modes generates effective long-range interactions
\cite{nelson87,aronovitz89}.

This mechanism provides the physical origin of the flat phase and of
its anomalous elasticity, to which we now turn.

\section{Pure polymerized membranes: the reference fixed point}
\label{sec:pure}

\subsection{Anomalous elasticity}

The flat phase is characterized by anomalous correlation functions
\cite{aronovitz88,guitter88,guitter89,aronovitz89,ledoussal92}.
The scaling form of the flexural  correlations is given by 
\begin{equation}
    G_{hh}(q)
    \equiv
    \left\langle
    \mathbf{h}(\mathbf{q})\cdot
    \mathbf{h}(-\mathbf{q})
    \right\rangle
    \sim
    q^{-(4-\eta)},
    \label{eq:ghh}
\end{equation}
while the phonon correlations scale as
\begin{equation}
    G_{uu}(q)
    \sim
    q^{-(2+\eta_u)}.
    \label{eq:guu}
\end{equation}
Residual rotational invariance leads to the Ward identity
\cite{aronovitz88,guitter88,aronovitz89,guitter89,ledoussal92} 
\begin{equation}
    \eta_u+2\eta=4-D.
    \label{eq:ward}
\end{equation}

Equivalently, one can express the anomalous elasticity through
scale-dependent elastic constants,
\begin{equation}
    \kappa_{\mathrm{eff}}(q)
    \sim q^{-\eta},
    \qquad
    \mu_{\mathrm{eff}}(q),\lambda_{\mathrm{eff}}(q)
    \sim q^{\eta_u}.
    \label{eq:elastic-scaling}
\end{equation}
From (\ref{eq:elastic-scaling}) one sees that thermal fluctuations make the membrane increasingly difficult to
bend while simultaneously softening its in-plane elastic response.
This counterintuitive fluctuation-induced stiffening is the defining
property of the flat phase \cite{nelson87,aronovitz88,aronovitz89}.

For the physical case $D=2$, Eq.~\eqref{eq:ward} becomes
\begin{equation}
    \eta_u=2-2\eta.
\end{equation}
Since $\eta>0$, the growth of $\kappa_{\mathrm{eff}}$ regularizes the
angular fluctuations:
\begin{equation}
    \left\langle\theta^2\right\rangle
    \sim
    \int d^2q\,
    \frac{1}{\kappa_{\mathrm{eff}}(q)q^2}
    \sim
    \int_0 dq\,q^{\eta-1},
\end{equation}
which is infrared convergent for $\eta>0$. Long-range orientational
order is therefore self-consistently stabilized by thermal
fluctuations themselves \cite{nelson87}.

\subsection{Weak-coupling expansion around $D=4$}

The upper critical dimension of the flat-phase theory is $D=4$
\cite{aronovitz88,guitter89,aronovitz89}.
Consequently, a natural perturbative parameter is 
\begin{equation}
\epsilon=4-D .
\end{equation}

The Lam\'e coefficients have canonical dimension $\epsilon$ and the RG
flow can be calculated as a weak-coupling expansion in $\lambda$,
$\mu$, and $\epsilon$.

The seminal perturbative calculations of Aronovitz and Lubensky and
of Guitter, David, Leibler and Peliti revealed a non-Gaussian fixed
point controlling the infrared physics of the flat phase
\cite{aronovitz88,guitter89,aronovitz89}. We shall denote this fixed point by
$\Pfour$, following the notation that will later be convenient in the
disordered problem.

At $\Pfour$, the bending rigidity diverges and the elastic moduli
vanish algebraically according to
Eq.~\eqref{eq:elastic-scaling}. Thus $\Pfour$ is the fixed point
responsible for the anomalous elasticity and long-range orientational
order of the flat phase \cite{aronovitz88,guitter89,aronovitz89}.

At one-loop order, however, quantitative extrapolation to the physical
dimension is not very accurate. For $D=2$ and $d=3$ the leading
$\epsilon$ expansion gives approximately
\cite{aronovitz88}
\begin{equation}
    \eta_{\mathrm{1L}}\simeq 0.96,
\end{equation}
whereas numerical simulations and more elaborate analytical approaches
typically yield values closer to $0.8$--$0.9$ \cite{ledoussal92,kownacki09,los09,troster13,troster15}. 
For recent Fourier Monte Carlo results for arbitrary numbers of
flexural modes $d_c$, see Ref.~\cite{ivanov24}. This discrepancy is
not surprising since the physical value corresponds to
$\epsilon=2$, well outside the immediate vicinity of $D=4$.

\subsection{Large-$d$ expansion and self-consistent screening}

An alternative route consists in exploiting the embedding dimension,
or equivalently the codimension $d_c=d-D$, as an expansion parameter.
The large-$d$ approach and its self-consistent implementation in the
SCSA resum infinite classes of diagrams and yield a much improved description of the physical membrane
\cite{ledoussal92,gazit09,zakharchenko10,roldan11,ledoussal18}.

Within the SCSA, the renormalization of the bending rigidity is
determined self-consistently by the screened elastic interactions. In
the infrared one finds again
\begin{equation}
    \kappa_{\mathrm{eff}}(q)\sim q^{-\eta},
\end{equation}
confirming that the anomalous rigidity is not an artifact of the $\epsilon$ expansion \cite{ledoussal92}.

For $D=2$ and $d=3$, the SCSA gives \cite{ledoussal92,ledoussal18}
\begin{equation}
    \eta_{\mathrm{SCSA}}\simeq0.821,
\end{equation}
which lies much closer to numerical estimates than the leading
$\epsilon$ expansion. This agreement was a major indication that the
flat-phase fixed point provides the correct effective description of
real crystalline membranes. Recent four-loop calculations have also
shed light on the remarkable accuracy of the SCSA at this fixed point.
By introducing a parameter that continuously switches on vertex
corrections, Métayer and Teber \cite{metayerteber25} showed that, in the physical case,
the anomalous dimension at $\Pfour$ can be written as
\begin{equation}
  \eta(\Pfour)=0.8667+0.0003\,V,
\end{equation}
where $V=0$ corresponds to neglecting vertex corrections, as in the
SCSA, while $V=1$ restores the full perturbative result. Vertex
corrections therefore produce only a minute effect at $\Pfour$, of
order $0.03\%$, as a consequence of strong cancellations among
diagrams. This provides a posteriori a simple explanation for the
quantitative success of the SCSA in the pure flat phase
\cite{metayerteber25}.

Nevertheless, going systematically beyond leading order is technically difficult, see \cite{gazit09}. The theory involves both phonon and flexural fields,
several elastic couplings and derivative interactions, producing
momentum-dependent vertices. These complications provided an important
motivation for developing a functional, nonperturbative RG
description \cite{kownacki09,braghin10,hasselmann11,essafi14,delzescaux26}.

\subsection{Nonperturbative renormalization group}

A complementary description is provided by the effective average
action $\Gamma_k$.  It interpolates between the microscopic action at
an ultraviolet scale $\Lambda$ and the full Gibbs free energy in the
infrared,
\begin{equation}
  \Gamma_{k=\Lambda}\simeq S,
  \qquad
  \Gamma_{k=0}=\Gamma.
\end{equation}
Its exact evolution with the running scale $k$  is given by the Wetterich equation
\cite{wetterich93},
\begin{equation}
  \partial_t\Gamma_k[\rr]
  =
  \frac{1}{2}
  \mathrm{Tr}
  \left[
    \left(\Gamma_k^{(2)}[\rr]+R_k\right)^{-1}
    \partial_tR_k
  \right],
  \qquad
  t=\ln(k/\Lambda).
  \label{eq:wetterich}
\end{equation}
where $R_k(\q)$ is a $\q$-dependent cutoff function.  
Although Eq.~\eqref{eq:wetterich} has a one-loop form, it contains the
full field-dependent inverse propagator.  It is therefore not an
expansion in the elastic couplings, in $\epsilon$, or in $1/d$
\cite{wetterich93,berges02,dupuis21}. 

For polymerized membranes, a rotationally invariant derivative
expansion can be organized directly in terms of the embedding field $\rr$. 
A minimal ansatz takes the schematic form \cite{kownacki09}
\begin{align}
  \Gamma_k[\rr]
  =
  \int\dd^Dx\Bigg\{
  &
  \frac{Z_k}{2}(\Delta\rr)^2 +{\lambda_k\over 8} 
  \left(
    \partial_a\rr\cdot\partial_a\rr
    -
    D\zeta_k^2
  \right)^2 + 
{\mu_k\over 4} 
  \left(
    \partial_a\rr\cdot\partial_b\rr
    -
    \zeta_k^2\delta_{ab}
  \right)^2 +\cdots
  \Bigg\}.
  \label{eq:nprg-pure}
\end{align}
where $a,b=1\dots D$. Here $Z_k$, $\lambda_k$, $\mu_k$ and $\zeta_k$ are running quantities and the
ellipsis denotes operators neglected within the truncation. See \cite{delzescaux26} for a review and an  extension of the NPRG approach to graphene bilayers. 

A key advantage of this formulation is that the same effective action can describe both the crumpling transition and the flat phase \cite{kownacki09}.
The crumpled regime is associated with a vanishing order parameter, typically a vanishing stretching factor $\zeta_k$,  whereas the deep flat phase corresponds to a running stretching factor $\zeta_k$ staying finite. The resulting RG equations reproduce the
known $\epsilon$ and large-$d$ limits while remaining directly
applicable to $D=2$ and $d=3$
\cite{kownacki09}.

For the physical case the leading NPRG treatment gives \cite{kownacki09,braghin10,essafi14}
\begin{equation}
  \eta_{\mathrm{NPRG}}\simeq0.849,
  \label{eq:eta-pure-nprg}
\end{equation}
in close agreement with numerical simulations of graphene \cite{los09,los16} 
 and with other nonperturbative estimates.  An important feature of the
nonperturbative treatment of the flat phase of membranes is the stability of this number against
extensions of the truncation \cite{braghin10,hasselmann11,essafi14}.

Recent high-order perturbative calculations provide an instructive
comparison.  For $D=2,d=3$, the successive estimates quoted in the
higher-loop analyses are \cite{aronovitz88,coquand20a,metayer22,pikelner22,metayer24,metayerteber25}, see also \cite{mauri20} for a closely related model:
\begin{equation}
  \eta_{\mathrm{1L}}=0.960,\qquad
  \eta_{\mathrm{2L}}\simeq0.9139,\qquad
  \eta_{\mathrm{3L}}\simeq0.8872,\qquad
  \eta_{\mathrm{4L}}\simeq0.8670,
  \label{eq:pure-loop-series}
\end{equation}
and show an apparent convergence towards the nonperturbative value $\eta_{\mathrm{NPRG}}\simeq0.85$
\cite{kownacki09,essafi14}.
This convergence is particularly significant for the study of quenched
disorder developed below: before addressing the disordered theory, it
provides strong evidence that perturbative and nonperturbative
approaches describe the same underlying infrared physics of the
pure flat phase. 

It is particularly instructive to compare the four-loop perturbative series with the NPRG result reexpanded around \(D=4\)
\cite{pikelner22,metayer24,metayerteber25}.    After translating to our convention, the perturbative four-loop series at the
physical codimension reads \cite{pikelner22,metayer24,metayerteber25}
\begin{equation}
  \eta_{\mathrm{4L}}
  =
  0.4800\,\eps
  -0.01152\,\eps^2
  -0.00334\,\eps^3
  -0.00126\,\eps^4
  +O(\eps^5),
\end{equation}
while the NPRG result, reexpanded to the same order around $D=4$, is
\cite{metayerteber25}
\begin{equation}
  \eta_{\mathrm{NPRG,reexp}}
  =
  0.4800\,\eps
  -0.00918\,\eps^2
  -0.00333\,\eps^3
  -0.00111\,\eps^4
  +O(\eps^5).
\end{equation}

The numerical agreement therefore persists through $O(\eps^4)$:
the leading coefficient is identical and the higher-order
coefficients remain remarkably close. The agreement is not,
however, term-by-term at the analytical level, since the NPRG
expansion does not reproduce the transcendental structures
involving Riemann zeta values that occur in the exact perturbative
calculation. The comparison nevertheless provides a stringent
quantitative benchmark of the NPRG truncation
\cite{metayerteber25}.

\begin{keyresult}
The fixed point $\Pfour$ is the reference infrared fixed point for the
rest of the review.  It controls the anomalous elasticity of the pure
flat phase and remains the attractor of weakly disordered membranes at
sufficiently high temperature
\cite{aronovitz88,guitter89,nelson91,radzihovsky91b,coquand18}.
\end{keyresult}

\section{Metric and curvature disorder}
\label{sec:disorder-model}

\subsection{Disordered elastic action}

Quenched disorder can be introduced directly in the geometrical
elastic theory.  We consider a short-range random curvature field
$\mathbf{c}(\x)$ and a short-range random stress field
$\sigma_{ab}(\x)$.  A rotationally invariant long-distance action is
\cite{morse92a,morse92b,coquand18}
\begin{equation}
  S[\rr]
  =
  \int\dd^Dx
  \left[
    \frac{\kappa}{2}(\partial_a^2\rr)^2
    +
    \frac{\lambda}{2}u_{aa}^2
    +
    \mu u_{ab}^2
    -
    \mathbf{c}(\x)\cdot\partial_a^2\rr
    -
    \sigma_{ab}(\x)u_{ab}
  \right].
  \label{eq:disordered-action}
\end{equation}
The curvature field $\mathbf{c}$ generates a local preferred spontaneous
curvature, while the random stress $\sigma_{ab}$ generates a local deformation of
the preferred metric \cite{morse92a,morse92b,coquand18}.

The disorder fields are taken to be Gaussian, short ranged, and of
zero mean, with variances \cite{morse92a,morse92b,coquand18}
\begin{equation}
  \big[
  c_i(\x)c_j(\x')
  \big]
  =
  \Delta_\kappa\,
  \delta_{ij}\,
  \delta^{(D)}(\x-\x'),
  \label{eq:curv-disorder}
\end{equation}
where $i,j=1\dots d$ and
\begin{equation}
  \big[
  \sigma_{ab}(\x)\sigma_{cd}(\x')
  \big]
  =
  \left(
    \Delta_\lambda\delta_{ab}\delta_{cd}
    +
    2\Delta_\mu I_{abcd}
  \right)
  \delta^{(D)}(\x-\x'),
  \label{eq:metric-disorder}
\end{equation}
where
\begin{equation}
  I_{abcd}
  =
  \frac{1}{2}
  \left(
    \delta_{ac}\delta_{bd}
    +
    \delta_{ad}\delta_{bc}
  \right).
\end{equation}
The variances $\Delta_\kappa$ and
$(\Delta_\lambda,\Delta_\mu)$ characterize curvature and metric
disorders, respectively.  Positivity requires
$\Delta_\kappa>0$, $\Delta_\mu>0$, and
$\Delta_\lambda+2\Delta_\mu/D>0$.

In the pure system, the flat reference configuration introduced above
satisfies $\partial_a\rr_0=\zeta\,\mathbf e_a$ and
$g^0_{ab}=\zeta^2\delta_{ab}$. Quenched disorder modifies this picture by making the preferred local
geometry spatially inhomogeneous. The random stress field
$\sigma_{ab}(\x)$ represents frozen local fluctuations of the preferred
metric, whereas the random field $\mathbf c(\x)$ imposes a frozen local
preferred curvature. Thus, unlike in the pure membrane, the geometry
toward which the membrane locally tends to relax varies from point to
point. 

The two disorder sources play physically different roles.  In
self-consistent descriptions curvature disorder can enhance the
effective rigidity, whereas metric disorder tends to soften the
membrane \cite{nelson91,radzihovsky91b,morse92a,morse92b}.  The long-distance problem is therefore intrinsically a
competition not only between temperature and disorder, but also
between the two geometrical manifestations of disorder.

\subsection{Replica formulation}

Since the disorder is quenched, the relevant free energy contains
$[\ln Z]$, where square brackets denote disorder averaging.  The
replica construction used for this membrane problem
\cite{morse92a,morse92b,coquand18} relies on the identity
\begin{equation}
  [\ln Z]
  =
  \lim_{n\to0}
  \frac{[Z^n]-1}{n}
  \label{eq:replica}
\end{equation}
which converts the disorder average into an effective interacting
$n$-replica theory.  In a flat-phase formulation based on the phonon
and flexural fields, the replicated couplings may be written
\begin{equation}
  \kappa_{\alpha\beta}
  =
 \kappa\,\delta_{\alpha\beta}
  -
  \Delta_\kappa J_{\alpha\beta},
\end{equation}
with analogous expressions for the Lam\'e couplings,
\begin{equation}
  \mu_{\alpha\beta}
  =
  \mu\,\delta_{\alpha\beta}
  -
  \Delta_\mu J_{\alpha\beta},
  \qquad
 \lambda_{\alpha\beta}
  =
 \lambda\,\delta_{\alpha\beta}
  -
  \Delta_\lambda J_{\alpha\beta},
  \label{eq:replica-couplings}
\end{equation}
where $J_{\alpha\beta}=1$ for all replica indices.

This representation makes clear why the disordered theory is
technically demanding.  Even the pure membrane already contains
derivative interactions and several elastic couplings.  Quenched
disorder adds replica structure and three disorder variances, while
the zero-temperature regime requires combinations of couplings that
remain finite when the curvature disorder diverges.

\subsection{Connected and disconnected correlations}

The presence of quenched disorder requires two distinct correlation
functions \cite{ledoussal92,morse92a,morse92b,coquand18,ledoussal18}.  Writing
\begin{equation}
  \delta {\bf h}(\q)
  =
  {\bf h}(\q)-\langle {\bf h}(\q)\rangle,
\end{equation}
the total height correlation function decomposes as
\begin{equation}
  G_{hh}(q)
  =
  T\chi(q)+C(q),
  \label{eq:decomposition-correlator}
\end{equation}
where
\begin{equation}
  T\chi(q)
  =
  \big[
    \langle
      \delta {\bf h}(\q). \delta {\bf h}(-\q)
    \rangle
  \big]
\end{equation}
is the connected, thermal contribution, and
\begin{equation}
  C(q)
  =
  \big[
    \langle {\bf h}(\q)\rangle
    .\langle {\bf h} (-\q)\rangle
  \big]
\end{equation}
is the disconnected, disorder-induced contribution.  At low momentum
they scale as
\begin{equation}
  \chi(q)\sim q^{-(4-\eta)},
  \qquad
  C(q)\sim q^{-(4-\eta')}.
  \label{eq:eta-etaprime}
\end{equation}
Rotational invariance gives separate Ward identities for thermal and
disorder-induced correlations,
\begin{equation}
  \eta_u+2\eta=4-D,
  \qquad
  \eta'_u+2\eta'=4-D.
\end{equation}

It is useful to introduce
\begin{equation}
  \phi=\eta'-\eta.
  \label{eq:phi}
\end{equation}
The sign of $\phi$ characterizes which contribution dominates the
infrared behavior \cite{morse92a,morse92b,coquand18,metayer22d}.  With the conventions used in the perturbative
analysis, $\phi>0$ corresponds to thermal dominance,
$\phi<0$ to disorder dominance, and $\phi=0$ to a marginal situation
in which the two sectors scale identically.

Finally, the corresponding roughness exponents, which characterize the membrane morphology, are
\begin{equation}
  \zeta
  =
  \frac{4-D-\eta}{2},
  \qquad
  \zeta'
  =
  \frac{4-D-\eta'}{2}.
  \label{eq:roughness}
\end{equation}
These relations will be central when comparing the RG predictions
with the experiments on partially polymerized lipid membranes.

\section{The early perturbative picture}
\label{sec:historical}

\subsection{Metric disorder}

The first systematic RG studies focused on disorder of the preferred
metric
\cite{nelson91,radzihovsky91b}.
Near the upper critical dimension, the disorder-free fixed point
$\Pfour$ remains stable at finite temperature.  In RG language the
metric disorder is irrelevant and the flow returns to the same
infrared universality class as a pure polymerized membrane. 

At very low temperature, however, disorder becomes increasingly
important.  This observation motivated the possibility of a
spin-glass-like phase in which static geometrical fluctuations would
dominate \cite{nelson91,radzihovsky91b}.  Large-$d$ and self-consistent analyses also suggested an
instability toward a glassy regime and introduced an
Edwards--Anderson-type characterization of frozen deformations \cite{ledoussal92,ledoussal18}.

\subsection{Adding curvature disorder}

Morse, Lubensky, and Grest \cite{morse92a} and subsequently Morse and
Lubensky \cite{morse92b} extended the weak-coupling analysis to the
combined metric--curvature disorder problem. Related effects of quenched
curvature disorder were also investigated by Bensimon, Mukamel, and
Peliti \cite{bensimon92}. In the metric--curvature model of
Refs.~\cite{morse92a,morse92b}, the one-loop analysis in
$D=4-\epsilon$ yields, in addition to the Gaussian point, two important
nontrivial fixed points.

The first is the familiar disorder-free point $\Pfour$.  At one loop,
using $\dc=d-D$,
\begin{equation}
  \mu_4
  =
  \frac{96\pi^2}{24+\dc}\,\eps,
  \qquad
  \Delta_{\mu,4}
  =
  \Delta_{\kappa,4}
  =
  0,
\end{equation}
and
\begin{equation}
  \eta_4
  =
  \frac{12}{24+\dc}\,\eps.
\end{equation}
This fixed point is attractive and controls the infrared behavior of
weakly disordered membranes at finite temperature.

The second is the disorder-dominated point $\Pfive$, located at
vanishing temperature and infinite curvature disorder.  To describe
this limit it is convenient to introduce the  combinations  
\begin{equation}
  g_\mu=\mu\Delta_\kappa,
  \qquad
  g_\lambda=\lambda\Delta_\kappa.
  \label{combi}
\end{equation}
Indeed, in the rescaled variables used here, the Lamé coefficients
scale as $T$, the metric-disorder variances as $T^0$, and the
curvature-disorder variance $\Delta_{\kappa}$ as $T^{-1}$.
Consequently, $g_\mu=\mu\Delta_\kappa$ and
$g_\lambda=\lambda\Delta_\kappa$ remain finite in the
zero-temperature limit, while $\mu\to0$; see Ref.~\cite{coquand18}.

At one loop, the coordinates needed below are
\begin{equation}
  \Delta_{\mu,5}
  =
  \frac{24\pi^2}{6+\dc}\,\eps,
  \qquad
  g_{\mu,5}
  =
  \frac{48\pi^2}{6+\dc}\,\eps,
\end{equation}
and
\begin{equation}
  \eta_5=\eta'_5
  =
  \frac{3}{6+\dc}\,\eps.
  \label{eq:P5-one-loop}
\end{equation}
Thus $\phi_5=0$ at this order.  An analysis of nonlinear corrections to
the flow shows that $\Pfive$ is marginally unstable with respect to
temperature \cite{morse92b}.  The one-loop conclusion is therefore that $\Pfive$ may
control a broad low-temperature scaling regime, but not an entire
stable thermodynamic phase.

The same qualitative structure was obtained within SCSA-based
treatments: $\Pfour$ is stable, a zero-temperature disordered point
exists, but the finite-temperature critical fixed point required for
a transition into a glassy phase is {\sl absent} \cite{ledoussal18}.

\begin{openproblem}
The experiments on partially polymerized membranes suggested a
wrinkling transition, whereas the leading weak-coupling and SCSA
descriptions did not contain a critical fixed point capable of
controlling such a transition.  This mismatch is the central
historical puzzle of the disordered-membrane problem
\cite{mutz91,chaieb06,morse92a,morse92b,coquand18}.
\end{openproblem}

\section{Nonperturbative RG and the emergence of a glassy phase}
\label{sec:nprg-disorder}

\subsection{Effective average action for the replicated theory}

The NPRG treatment starts by averaging the disorder through the
replica formalism and constructing a scale-dependent effective action
$\Gamma_k[\{\rr\}]$ \cite{coquand18}.  The flow is again governed by the exact
Wetterich equation,
\begin{equation}
  \partial_t\Gamma_k[\rr]
  =
  \frac{1}{2}
  \mathrm{Tr}
  \left\{
    \partial_tR_k
    \left(
      \Gamma_k^{(2)}[\rr]+R_k
    \right)^{-1}
  \right\}.
  \label{eq:wetterich-disorder}
\end{equation}
The trace now includes momentum, embedding-space, and replica
indices.

A rotationally invariant truncation is constructed directly in terms
of the embedding fields $\{\rr^{\alpha}\}$.  Schematically, it contains the running
bending term, the elastic invariants, and inter-replica contributions
associated with $\Delta_{\kappa,k}$,
$\Delta_{\lambda,k}$, and $\Delta_{\mu,k}$.  The crucial point is that
the flow retains nonpolynomial dependence on the disorder variables
and on the couplings.  When expanded near $D=4$ it reproduces the
one-loop weak-coupling equations, but away from $D=4$ is not restricted to a finite-order polynomial expansion. 

The running field renormalization and curvature-disorder variance
define two anomalous dimensions.  At a fixed point they coincide with
the exponents $\eta$ and $\eta'$ appearing in
Eq.~\eqref{eq:eta-etaprime}.

\subsection{Three nontrivial fixed points}

The nonperturbative flow in the physical case $D=2,d=3$ displays three
nontrivial fixed points of direct physical relevance \cite{coquand18}.

\paragraph{The thermal fixed point $\Pfour$.}
The point $\Pfour$ lies at finite temperature and vanishing disorder.
It is the continuation of the pure flat-phase fixed point and is fully
attractive within the high-temperature basin.  A weak amount of
quenched disorder is therefore asymptotically irrelevant in this
phase.

\paragraph{The disorder-dominated fixed point $\Pfive$.}
The point $\Pfive$ lies at vanishing temperature. In the original
variables, the curvature-disorder variance $\Delta_\kappa$ diverges,
whereas the rescaled disorder combinations introduced in
Eq.~\eqref{combi} remain finite. In contrast with the leading perturbative
result, $\Pfive$ is fully attractive within the NPRG flow.

\paragraph{The critical point $\Pc$.}
A third fixed point $P_c$ lies on the critical separatrix separating the basins of attraction of $\Pfour$ and $\Pfive$.  It occurs at
finite temperature and finite disorder and has one relevant
temperature-like direction.  It therefore controls a continuous
transition between the two basins of attraction.

\begin{figure}[t]
\includegraphics[scale = 1]{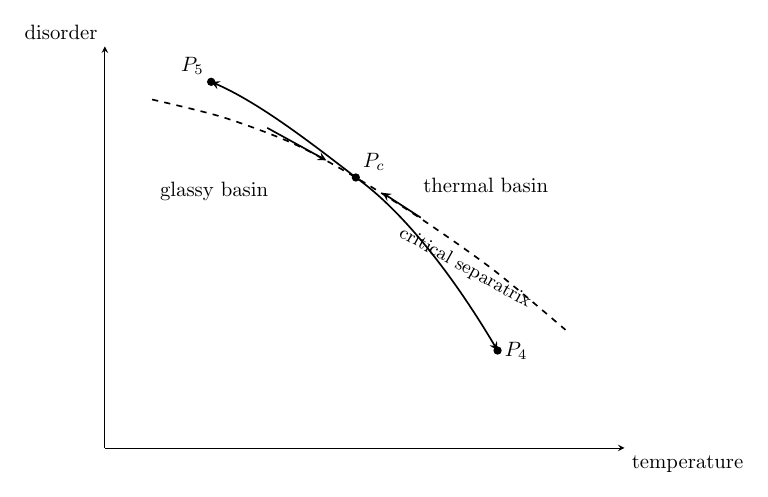}
\caption{Schematic RG topology of the quenched metric--curvature
disorder problem. The dashed curve represents the critical separatrix,
i.e. the stable manifold of the finite-temperature fixed point $\Pc$;
the arrows drawn along it indicate that RG trajectories on the
separatrix flow toward $\Pc$. The solid trajectories leaving $\Pc$
represent the flow along its relevant direction, toward either the
thermal fixed point $\Pfour$ or the disorder-dominated fixed point
$\Pfive$. The two sides of the separatrix define the thermal and glassy
basins of attraction, respectively. The temperature and disorder axes
are schematic, and the drawing does not represent the actual
multidimensional coupling space.}
\label{fig:flowpicture}
\end{figure}

The topology of the flow can be represented schematically as
\begin{equation}
  \boxed{
  \Pfour
  \quad\longleftarrow\quad
  \Pc
  \quad\longrightarrow\quad
  \Pfive
  }.
  \label{eq:flow-topology}
\end{equation}
This should not be read as a one-dimensional flow; it is a symbolic
representation of the two basins separated by the critical
separatrix. A more realistic illustration is given in Fig. \ref{fig:flowpicture}. 

\subsection{Why $\Pc$ is invisible at one loop}

Close to $D=4$, the nonperturbative solution explains why the new
critical point was missed in the original perturbative calculation
\cite{coquand18}.
To leading nontrivial order,
\begin{equation}
  \mu_c
  =
  \frac{4\pi^2(27+5\dc)}
  {15(6+\dc)^2}\,
  \eps^2,
  \label{eq:muc-nprg}
\end{equation}
while the other leading coordinates coincide with those of $\Pfive$,
\begin{equation}
  \Delta_{\mu,c}
  =
  \frac{24\pi^2}{6+\dc}\,\eps,
  \qquad
  g_{\mu,c}
  =
  \frac{48\pi^2}{6+\dc}\,\eps.
\end{equation}
Moreover,
\begin{equation}
  \eta_c=\eta'_c
  =
  \frac{3}{6+\dc}\,\eps
\end{equation}
at leading order.  The point $\Pc$ is therefore degenerate with
$\Pfive$ at order $\epsilon$ and separates from it only at higher
order, notably through the nonzero temperature coordinate
$\mu_c=O(\epsilon^2)$ and the relevant eigenvalue
\begin{equation}
  y_c
  =
  \frac{(27+5\dc)\dc}
  {20(6+\dc)^3}\,
  \eps^2.
\end{equation}
A one-loop expansion cannot resolve this splitting.

\subsection{Physical exponents}

For $D=2,d=3$, the NPRG gives values near \cite{coquand18}
\begin{equation}
  \eta_5\simeq0.449,\qquad
  \eta'_5\simeq0.277,\qquad
  \phi_5\simeq-0.172,
  \label{eq:P5-nprg-old}
\end{equation}
for the disorder fixed point, and
\begin{equation}
  \eta_c=\eta'_c\simeq0.492
  \label{eq:Pc-nprg}
\end{equation}
for the critical point.  In the refined analysis used for comparison
with experiment the corresponding values are \cite{coquand20}
\begin{equation}
  \eta_5=0.448(2),\qquad
  \eta'_5=0.275(2),
  \qquad
  \eta_c=\eta'_c=0.490(2).
  \label{eq:nprg-refined}
\end{equation}
The negative value of $\phi_5$ means that disorder fluctuations
dominate the infrared scaling at $\Pfive$, whereas the equality
$\eta_c=\eta'_c$ characterizes the critical point
\cite{coquand18,coquand20}.

\begin{keyresult}
Within the NPRG, the qualitative change relative to the one-loop
picture is not merely the numerical value of an exponent.  It is a
change in the topology of the RG flow: the appearance of $\Pc$ makes
$\Pfive$ fully attractive and converts the disorder-dominated regime
into a genuine low-temperature phase \cite{coquand18}.
\end{keyresult}

\section{Higher-order perturbation theory}
\label{sec:higher-loop}

\subsection{Motivation}

The discovery of $\Pc$ within the NPRG immediately raised a natural
question: could the new fixed point be an artifact of the
nonperturbative truncation?  The issue was sharpened by the absence of
$\Pc$ in the one-loop weak-coupling analysis and in the available
SCSA treatment.  A perturbative calculation beyond leading order was
therefore an important independent test.

A two-loop calculation in $D=4-\epsilon$ identified a finite-temperature, finite-disorder $\Pc$-like fixed point
\cite{coquand20b}.  A later three-loop analysis corrected the
curvature-disorder RG function and determined the fixed-point
quantities through $O(\epsilon^3)$
\cite{metayer22d}.   The latter calculation was
performed in a modified minimal-subtraction scheme for the massless
theory.  The diagrammatic problem is technically demanding because the
vertices are momentum dependent and every graph carries a nontrivial
replica structure.

\subsection{Two-loop structure}

At two loops, $\Pfive$ remains a zero-temperature point with
$\mu=0$, while the $\Pc$-like solution acquires a temperature
coordinate of order $\epsilon^2$ \cite{coquand20b,metayer22d},
\begin{equation}
  \mu_c
  =
  -\frac{8\pi^2}{(6+\dc)^2}\,\epsilon^2
  +O(\epsilon^3).
  \label{eq:muc-pert}
\end{equation}
Its other leading coordinates coincide with those of $\Pfive$.
For the physical codimension $\dc=1$, the anomalous dimensions at
$\Pfive$ are
\begin{align}
  \eta_5
  &=
  0.42857\,\epsilon
  -
  0.03779\,\epsilon^2
  +O(\epsilon^3),
  \\
  \eta'_5
  &=
  0.42857\,\epsilon
  -
  0.03341\,\epsilon^2
  +O(\epsilon^3),
  \label{eq:P5-2loop}
\end{align}
so that
\begin{equation}
  \phi_5=0.00437\,\epsilon^2>0.
\end{equation}
According to the convention introduced above, the positive value of
$\phi_5$ signals the growing importance of thermal fluctuations and is
therefore consistent with the perturbative instability of $\Pfive$
against temperature. 
At the critical point,
\begin{equation}
  \eta_c=\eta'_c
  =
  0.42857\,\epsilon
  -
  0.03695\,\epsilon^2
  +O(\epsilon^3).
  \label{eq:Pc-2loop}
\end{equation}

The existence of a separate $\Pc$-like solution is the key
qualitative result \cite{coquand20b,metayer22d}.  At the same time, two difficulties appear.
First, the coordinate $\mu_c$ is negative at the leading order at
which $\Pc$ separates from $\Pfive$, apparently violating the
mean-field stability condition on the shear modulus.  Second, the resulting stability structure differs qualitatively from
the NPRG one.

In the exact theory, fixed-point coordinates are scheme- and parametrization-dependent, whereas critical exponents and stability eigenvalues are universal. At finite perturbative order, however, residual scheme dependence remains in their estimates. Similar apparent excursions of physical fixed points outside their mean-field stability domains occur in the higher-loop pure-membrane problem, even though the corresponding universal exponents remain well behaved \cite{metayer22,pikelner22}.

\subsection{Three-loop anomalous dimensions}

At three loops the qualitative perturbative picture is unchanged:
the flow contains $\Pfour$, $\Pfive$, and a finite-temperature,
finite-disorder $\Pc$-like solution \cite{metayer22d}.  For $\dc=1$ the anomalous
dimensions at $\Pfive$ are
\begin{align}
  \eta_5
  &=
  0.42857\,\epsilon
  -
  0.03779\,\epsilon^2
  -
  0.01205\,\epsilon^3
  +O(\epsilon^4),
  \label{eq:eta5-3loop}
  \\
  \eta'_5
  &=
  0.42857\,\epsilon
  -
  0.03341\,\epsilon^2
  -
  0.00964\,\epsilon^3
  +O(\epsilon^4),
  \label{eq:etap5-3loop}
\end{align}
while at $\Pc$,
\begin{equation}
  \eta_c=\eta'_c
  =
  0.42857\,\epsilon
  -
  0.03695\,\epsilon^2
  -
  0.01191\,\epsilon^3
  +O(\epsilon^4).
  \label{eq:etac-3loop}
\end{equation}

The comparison with the NPRG reexpanded around $D=4$ is especially
revealing \cite{metayer22d}.  The latter gives
\begin{align}
  \eta_{5,\mathrm{NPRG,reexp}}
  &=
  0.42857\,\epsilon
  -
  0.03532\,\epsilon^2
  -
  0.01293\,\epsilon^3,
  \\
  \eta'_{5,\mathrm{NPRG,reexp}}
  &=
  0.42857\,\epsilon
  -
  0.03999\,\epsilon^2
  -
  0.01636\,\epsilon^3,
  \\
  \eta_{c,\mathrm{NPRG,reexp}}
  &=
  0.42857\,\epsilon
  -
  0.03621\,\epsilon^2
  -
  0.01318\,\epsilon^3.
  \label{eq:nprg-reexpanded-disorder}
\end{align}

At $P_c$, the perturbative and reexpanded NPRG coefficients are remarkably close in magnitude and display the same sign structure through the orders shown.

If one simply evaluates the direct perturbative series at
$\epsilon=2$, the successive estimates for the critical exponent are
\begin{equation}
  \eta_c^{(1)}=0.85714,\qquad
  \eta_c^{(2)}=0.70933,\qquad
  \eta_c^{(3)}=0.61402,
  \label{eq:eta-c-sequence}
\end{equation}
to be compared with the NPRG value $\eta_c\simeq0.492$.  Similarly,
for $\Pfive$ one obtains
\begin{equation}
  \eta_5^{(1)}=0.85714,\qquad
  \eta_5^{(2)}=0.70600,\qquad
  \eta_5^{(3)}=0.60962,
\end{equation}
to be compared with $\eta_5\simeq0.449$ from the NPRG.  The finite
series therefore move monotonically toward the nonperturbative
estimates, although the three-loop values remain quantitatively above
them in $D=2$.

The key result is not that the direct $\epsilon=2$ substitution has
already converged numerically, but that the perturbative and NPRG
expansions display the same structure and very similar coefficients,
while both contain a separate finite-temperature, finite-disorder
critical solution.

\subsection{What is confirmed, and what remains unresolved?}

The higher-loop perturbative calculations \cite{coquand20b,metayer22d} provide strong independent support for the existence of the $P_c$-like solution and for the anomalous dimensions predicted by the NPRG \cite{coquand18}. At the present finite order, however, they do not reproduce the full
NPRG stability topology: the perturbative $\Pc$ has an apparently
unphysical negative shear-modulus coordinate, and the stability of
$\Pc$ and $\Pfive$ is reversed.

The robust statement is that the critical solution and its universal anomalous dimensions are
seen by both approaches.  The precise finite-order fixed-point
coordinates and stability eigenvalues require further control, for
example through alternative renormalization schemes, higher orders,
or appropriate resummations.

\begin{physicalmessage}
The perturbative and nonperturbative approaches are most convincingly
compared through universal exponents, not through finite-order
fixed-point coordinates.  Their agreement on the existence and
anomalous dimensions of the $\Pc$ solution is strong; their agreement
on the complete stability topology is not yet exact.
\end{physicalmessage}

\section{Wrinkling transition in partially polymerized membranes}
\label{sec:experiment}

\subsection{Experimental phenomenology}

The experiments on partially polymerized phospholipid vesicles provide
the clearest available phenomenological test of the disordered-membrane
scenario
\cite{mutz91,chaieb06,chaieb08}.  The membranes are partially
polymerized by ultraviolet irradiation and, upon cooling, undergo a
reversible transformation between relatively smooth and strongly
wrinkled morphologies.

The experiments summarized in the theoretical analysis cover vesicles ranging in size from approximately $0.3$ to  $40\,\mu$m
\cite{mutz91,chaieb06,coquand20}.
After polymerization and reheating, cooling to temperatures in the range 
$18$--$22^\circ$C produces a spontaneous wrinkling transition.  The
surface morphology was characterized by techniques including small-angle neutron scattering, environmental scanning electron microscopy,
and atomic force microscopy.

The most useful quantity for comparison with the RG is the one-dimensional power spectrum of a height profile,
\begin{equation}
  P(k)=\langle\vert \tilde h({\bf k})\vert^2\rangle \sim k^{-\gamma}.
  \label{eq:power-spectrum}
\end{equation}
Equivalently, for $k\ne 0$, it has the same scaling as the Fourier transform of the mean-square height difference $\langle (h(\x)-h({\bf 0}))^2\rangle$. 

In this one-dimensional convention, the power exponent is related to the roughness exponent by
\begin{equation}
  \gamma=1+2\zeta.
  \label{eq:gamma-zeta-exp}
\end{equation}
Three distinct regimes were identified as the degree of
polymerization was varied \cite{chaieb06,coquand20}.

At low polymerization, roughly below $30\%$, the membranes display
large creases characteristic of the wrinkled state, with
\cite{chaieb06,coquand20}
\begin{equation}
  \gamma_{\mathrm{exp}}\simeq2.9\pm0.1,
  \qquad
  \zeta_{\mathrm{exp}}\simeq0.95\pm0.05.
  \label{eq:exp-low}
\end{equation}
At high polymerization, approximately $32$--$40\%$, the vesicles are
much smoother and one finds \cite{chaieb06,coquand20}
\begin{equation}
  \gamma_{\mathrm{exp}}\simeq2.00\pm0.06,
  \qquad
  \zeta_{\mathrm{exp}}\simeq0.50\pm0.03.
  \label{eq:exp-high}
\end{equation}
In the intermediate range, around $30$--$32\%$, the morphology is
critical and \cite{chaieb06,coquand20}
\begin{equation}
  \gamma_{\mathrm{exp}}\simeq2.51\pm0.03,
  \qquad
  \zeta_{\mathrm{exp}}\simeq0.75\pm0.02.
  \label{eq:exp-critical}
\end{equation}

The existence of three robust scaling regimes is precisely what makes
this system particularly informative from the RG point of view.

\subsection{Scaling predictions from the fixed points}

In the disordered theory, the long-distance behavior of the total
height correlation function
\begin{equation}
  G_{hh}(q)=T\chi(q)+C(q)
\end{equation}
is controlled by whichever contribution dominates at the relevant
fixed point. For the one-dimensional power spectrum used in the experiments,  Eqs.~\eqref{eq:roughness} and \eqref{eq:gamma-zeta-exp} imply, in general internal dimension $D$, 
\begin{equation}
  \gamma
  =
  5-D-\eta
\end{equation}
when the connected thermal contribution dominates, and
\begin{equation}
  \gamma
  =
  5-D-\eta'
\end{equation}
when the disconnected disorder contribution dominates.  In the
physical case $D=2$, these reduce to
\begin{equation}
  \gamma=3-\eta
  \qquad\text{or}\qquad
  \gamma=3-\eta'.
  \label{eq:gamma-eta}
\end{equation}

At $\Pfour$, the thermal fixed point, the refined NPRG value
\cite{coquand20}
\begin{equation}
  \eta_4=0.849(3)
\end{equation}
gives
\begin{equation}
  \zeta_4=0.575(2),
  \qquad
  \gamma_4=2.151(3).
  \label{eq:P4-gamma}
\end{equation}
This is close to the high-polymerization smooth regime, in which the
measured power exponent saturates near $2$ \cite{chaieb06,coquand20}.

At the critical point $\Pc$ \cite{coquand20},
\begin{equation}
  \eta_c=\eta'_c=0.490(2),
\end{equation}
hence
\begin{equation}
  \zeta_c=0.755(1),
  \qquad
  \gamma_c=2.510(2).
  \label{eq:Pc-gamma}
\end{equation}
The agreement with the experimental critical value
$\gamma=2.51\pm0.03$ is {\sl striking} \cite{chaieb06,coquand20}.

Finally, at the disorder-dominated point $\Pfive$
\cite{coquand20},
\begin{equation}
  \eta_5=0.448(2),
  \qquad
  \eta'_5=0.275(2).
\end{equation}
Because the disconnected contribution controls the long-distance
height fluctuations, the relevant roughness exponent is
\begin{equation}
  \zeta'_5=0.862(1),
\end{equation}
which yields
\begin{equation}
  \gamma_5=2.725(2).
  \label{eq:P5-gamma}
\end{equation}
This value lies close to the low-polymerization plateau observed in
the wrinkled regime \cite{chaieb06,coquand20}. The very-low-polymerization
data point with $\gamma\simeq3$ is not included in this plateau comparison;
it may indicate a crossover toward fluid-membrane behavior when the degree
of polymerization becomes too small for the polymerized elastic description
to remain appropriate \cite{coquand20}. The NPRG predictions and the
experimental results are summarized in Table~\ref{tab:experiment}.

\begin{table}[h]
\centering
\caption{Correspondence between RG fixed points and the three
experimental scaling regimes of partially polymerized lipid
membranes.  The theoretical numbers are the refined NPRG estimates
used in the comparison with experiment.}
\label{tab:experiment}
\vspace{0.5cm}
\begin{tabular}{lcccc}
\toprule
Regime & Fixed point & \hspace{0.5cm} Relevant exponent & $\gamma_{\rm th}$ &
$\gamma_{\rm exp}$\\
\midrule
Smooth, high polymerization
& \(\Pfour\) & \(\eta_4=0.849(3)\) & \(2.151(3)\) & \(2.00\pm0.06\)\\
Critical
& $\Pc$ & $\eta_c=\eta'_c=0.490(2)$ & $2.510(2)$ &
$2.51\pm0.03$\\
Wrinkled, low polymerization
& $\Pfive$ & $\eta'_5=0.275(2)$ & $2.725(2)$ &
$\simeq2.8$--$2.9$\\
\bottomrule
\end{tabular}
\end{table}

\subsection{Universal interpretation}

The importance of this comparison is not restricted to numerical
agreement.  Under the proposed identification, the three scaling behaviors are associated with fixed points or with RG flows controlled by fixed points and are therefore universal within the effective theory \cite{coquand20}.  The degree of polymerization
is a microscopic control parameter, whereas the exponents depend only
on the infrared fixed point reached by the flow.

The natural identification is therefore
\begin{equation}
  \Pfour
  \longleftrightarrow
  \text{smooth thermal flat phase},
\end{equation}
\begin{equation}
  \Pc
  \longleftrightarrow
  \text{wrinkling transition},
\end{equation}
and
\begin{equation}
  \Pfive
  \longleftrightarrow
  \text{wrinkled/glassy phase}.
\end{equation}

This correspondence provides phenomenological support for the nonperturbative three-fixed-point structure \cite{coquand20}.  Conversely, the
agreement with the experimental critical regime gives physical support
to the interpretation of $\Pc$ as the wrinkling critical point.

\section{A unified physical picture}
\label{sec:unified}

The results reviewed above can be organized around a simple sequence
of physical ideas.

For a pure polymerized membrane, thermal fluctuations generate
anharmonic interactions between bending and in-plane elasticity.  The
resulting infrared flow is toward $\Pfour$, where the bending rigidity
grows and the in-plane elastic moduli soften
\cite{nelson87,aronovitz88,guitter89,aronovitz89,ledoussal92}.  This anomalous elasticity
stabilizes orientational order and makes the flat phase possible.

Quenched disorder adds a second source of fluctuations.  Weak disorder
at finite temperature does not necessarily change the asymptotic
physics: trajectories in the thermal basin still flow to $\Pfour$
\cite{nelson91,radzihovsky91b,coquand18}.
In this sense the pure fixed point is robust.

At stronger disorder or lower temperature, however, the competition
changes character.  The disconnected correlation function becomes
increasingly important.  Within the NPRG flow the critical separatrix
is controlled by $\Pc$ \cite{coquand18}. Crossing this separatrix changes the infrared
attractor from $\Pfour$ to $\Pfive$ \cite{coquand18}.

The fixed point $\Pfive$ is qualitatively different from $\Pfour$.
Its effective temperature vanishes under coarse graining and the
long-distance morphology is controlled by frozen disorder
fluctuations \cite{morse92a,morse92b,coquand18}.  The terminology ``glassy'' used here refers precisely
to this disorder-dominated, frozen character.  It should not be
confused automatically with every structural notion of a glass in
microscopic materials, nor does the elastic theory by itself establish
all properties normally associated with spin glasses.

The resulting physical chain is
\begin{equation}
\begin{split}
  \text{pure membrane}
  &\rightsquigarrow
  \text{anomalous elasticity}
  \rightsquigarrow
  \Pfour,
  \\
  \text{quenched metric + curvature disorder}
  &\rightsquigarrow
  \text{competition of thermal and frozen fluctuations}
  \\
  &\rightsquigarrow \hspace{0.5cm} \Pfour
  \quad\longleftarrow\quad
  \Pc
  \quad\longrightarrow\quad
  \Pfive
\end{split}
\end{equation}

From this viewpoint, $P_c$  organizes the RG flow by separating two distinct infrared mechanisms of membrane stabilization: thermal anomalous elasticity on one side and disorder-controlled roughness on the other.

The perturbative and nonperturbative approaches illuminate
complementary parts of this picture.  The NPRG provides the global flow
and a direct treatment of $D=2$.  High-order perturbation theory tests
the structure close to $D=4$ with systematically computed loop
corrections.  Their agreement on the anomalous dimensions is consistent with their
probing the same fixed-point physics, while their disagreement on
finite-order stability properties identifies precisely where further
theoretical work is required \cite{metayer22d}.

Note finally that a  related finite-temperature transition between a thermally dominated flat phase and a disorder-dominated rippled phase has also been obtained
within a large-$d_c$ approach to flexible two-dimensional materials
\cite{saykin20b}. Although the precise relation between this transition
and the $\Pc$ fixed point discussed here remains to be clarified, this
result provides an independent indication that quenched disorder may
generate a finite-temperature transition between distinct flat regimes.

\section{Open problems and perspectives}
\label{sec:perspectives}

\subsection{Nature of the glassy phase}

The first open problem is the precise characterization of the
low-temperature phase.  The elastic RG shows that the infrared
behavior is controlled by quenched disorder and that the disconnected
correlation function dominates \cite{coquand18,coquand20}.  This motivates the use of the term “glassy” in a broad statistical-mechanical sense.  A more detailed
connection with conventional spin-glass concepts, including
Edwards--Anderson-type order parameters, remains desirable.

An important question is whether one can formulate observables that
distinguish unambiguously a genuine glassy membrane phase from a
strongly disordered but non-glassy elastic state.  Such observables
would be particularly useful for numerical simulations and
experiments.

\subsection{Perturbative stability and scheme dependence}

The three-loop calculation recovers $\Pc$ and produces anomalous
dimensions close to those of the NPRG, but the finite-order
fixed-point coordinates are problematic \cite{metayer22d}.  The negative perturbative
value of the shear-modulus coordinate at $\Pc$ and the reversed
stability of $\Pc$ and $\Pfive$ prevent a literal identification of
the complete perturbative flow with the NPRG flow.

Several routes are possible.  One may investigate alternative
regularization or renormalization schemes, extend the perturbative
series, or apply resummation techniques when sufficient orders are
available.  Since fixed-point coordinates are not universal, such
studies should focus primarily on universal exponents and stability
eigenvalues.

The SCSA also deserves renewed attention.  It provides very good
quantitative results for the pure membrane and for some properties of
$\Pfive$, yet the critical point $\Pc$ is absent in the available
SCSA treatment \cite{ledoussal92,coquand18}.  Understanding which partial resummations suppress or
retain the $\Pc$ solution would clarify the relation among the various
analytical methods.

\subsection{Improved NPRG truncations}

The NPRG results rely on a systematic  but finite truncation of the
effective average action.  In the pure flat phase, extensions of the
derivative expansion produce only small quantitative changes, which is
encouraging \cite{braghin10,essafi14}.  Nevertheless, the disordered problem contains more
couplings and a richer replica structure.  Systematic extensions of
the truncation would provide a direct test of the robustness of
$\Pc$, the stability exponent $y_c$, and the disorder exponent
$\eta'_5$.

\subsection{Numerical simulations}

A direct numerical test of the full $\Pfour$--$\Pc$--$\Pfive$
scenario remains highly desirable.  Simulations in which metric and
curvature disorder can be tuned independently would be especially
valuable.  Among the quantities to measure are the connected and
disconnected height correlations, the anomalous dimensions $\eta$ and
$\eta'$, the critical roughness exponent, and the flow of effective
elastic constants with system size.

Finite-size scaling near the transition could test whether the
wrinkling transition is continuous and determine the relevant
eigenvalue associated with $\Pc$.  Simulations could also determine
whether the disorder-dominated phase displays additional glassy
phenomena beyond the static roughness encoded in the present elastic
field theory.

\subsection{Partially polymerized membranes}

The quantitative agreement with the experiments of Chaieb and
collaborators is one of the strongest motivations for the present
scenario, but the experimental system is itself complex
\cite{chaieb06,chaieb08,coquand20}.  The degree
of polymerization is not identical to a single disorder variance in
the continuum action, and the relation between microscopic
polymerization patterns and effective metric/curvature disorder is not
known in detail.

Further measurements of both connected thermal fluctuations and frozen
sample-to-sample contributions would provide a sharper test of the
theoretical decomposition in Eq.~\eqref{eq:decomposition-correlator}.
It would also be useful to determine whether the transition can be
followed continuously in a sufficiently large scaling window to
extract crossover exponents in addition to the roughness exponent.

\subsection{Graphene and related two-dimensional materials}

Defective graphene and graphene-like crystalline monolayers are
natural candidates for extending the disordered-membrane scenario.
Vacancies and other lattice defects can modify the preferred metric,
while reconstructions into nonhexagonal rings can generate local
curvature \cite{amorim16,coquand18,coquand20}.  Thus both ingredients of the effective model can, in
principle, be present. Related effects of quenched disorder on the large-scale morphology of free-standing graphene, including disorder-induced rippling and
crumpling, have been investigated by Gornyi, Kachorovskii, and Mirlin
\cite{gornyi15}.

The extrapolation must nevertheless be made cautiously.  The elastic
model neglects electronic degrees of freedom and does not resolve the
microscopic chemistry of individual defects \cite{coquand18}.  Consequently, the
present theory predicts a possible universal long-distance elastic
scenario rather than a complete microscopic theory of defective
graphene.

A particularly interesting direction is the coupling between the
disorder-controlled membrane morphology and electronic, transport,
thermal, or optical properties.  If a wrinkling/glassy transition
occurs in an electronic crystalline membrane, the accompanying change
in large-scale geometry may feed back on the electronic sector.
Establishing this connection requires extending the theory beyond the
purely elastic degrees of freedom considered here.

\begin{openproblem}
The central experimental question for graphene-like materials is whether one can identify
a scale-invariant transition with the universal exponents associated
with $\Pc$ and a low-temperature regime governed by the
disorder-dominated fixed point $\Pfive$.
\end{openproblem}

\section{Conclusion}
\label{sec:conclusion}

Polymerized membranes display a hierarchy of fluctuation-induced
phenomena that make them a particularly instructive class of systems.  Even without quenched disorder, their flat
phase is nontrivial: the geometrical coupling between phonon and
flexural modes generates anomalous elasticity, with an effective
bending rigidity that diverges algebraically in the infrared.  The
associated fixed point $\Pfour$ is now supported by perturbative,
self-consistent, nonperturbative, and numerical approaches
\cite{aronovitz88,guitter89,ledoussal92,kownacki09,los09,metayer22,pikelner22}.

Quenched metric and curvature disorder produce a qualitatively richer
problem.  The early one-loop picture established the stability of
$\Pfour$ at finite temperature and identified the zero-temperature
disorder point $\Pfive$, but found the latter marginally unstable with
respect to temperature.  This framework therefore contained no
finite-temperature transition to a stable glassy phase
\cite{nelson91,radzihovsky91b,morse92a,morse92b}.

The NPRG changed this picture by revealing the finite-temperature,
finite-disorder fixed point $\Pc$.  Within the nonperturbative flow,
$\Pc$ separates the basin of the thermal fixed point $\Pfour$ from the
basin of the zero-temperature fixed point $\Pfive$.  The latter becomes
fully attractive and controls an entire disorder-dominated phase.  The
critical point is invisible at one loop because it separates from
$\Pfive$ only beyond leading order in $\epsilon$ \cite{coquand18}.

Higher-order perturbation theory provides an important independent
test.  A $\Pc$-like fixed point appears at two loops and persists at
three loops.  Most significantly, the three-loop anomalous dimensions
and the NPRG expressions reexpanded in $\epsilon$ have very similar
coefficients.  This strongly supports the identification of the
critical solution across the two approaches.  The remaining mismatch
in fixed-point coordinates and stability properties at finite
perturbative order should, however, be kept explicit: it is an open
technical issue rather than a resolved detail \cite{metayer22d}.

The connection with partially polymerized lipid membranes gives this
field-theoretical structure direct physical content.  The smooth,
critical, and wrinkled experimental regimes are naturally associated
with $\Pfour$, $\Pc$, and $\Pfive$, respectively.  In particular, the
theoretical critical value $\gamma_c=2.510(2)$ agrees with the measured
$\gamma=2.51\pm0.03$, while the high- and low-polymerization regimes
are also consistent with the corresponding fixed-point predictions.
Because these exponents are tied to RG fixed points, the interpretation
is one of universality rather than a microscopic fit to a specific
lipid system \cite{coquand20}.

The present picture therefore links three themes that were initially
developed separately: anomalous elasticity of pure polymerized
membranes, disorder-induced glassy behavior, and the experimentally
observed wrinkling transition.  The next stage is to test the global
RG scenario more directly through controlled simulations and new
experiments, and to determine whether related universal behavior can
be observed in defective graphene and other two-dimensional
crystalline materials.

\section*{Acknowledgements}

I would  like to thank S. Metayer  for  interesting  discussions  and insightful suggestions.


%

\end{document}